\documentclass[journal=jpccck,manuscript=article]{achemso}
\usepackage[version=3]{mhchem} 
\usepackage{amsmath,amssymb}
\usepackage{graphicx}
\usepackage{times}
\usepackage{fontenc}
\usepackage{epstopdf}
\usepackage{fancyhdr}
\usepackage{bm}
\usepackage{float}
\usepackage{color}
\usepackage{gensymb}
\usepackage{soul} 
\author{Babu Ram}
\affiliation[Materials Research Centre, Indian Institute of Science, Bangalore
  560012, India]
{Materials Research Centre, Indian Institute of Science, Bangalore
  560012, India}
\author{Aaditya Manjanath}
\affiliation[Materials Research Centre, Indian Institute of Science, Bangalore
  560012, India]
{Materials Research Centre, Indian Institute of Science, Bangalore
  560012, India}
\alsoaffiliation[Centre for Nano Science and Engineering, Indian Institute of Science, Bangalore
  560012, India]
{Centre for Nano Science and Engineering, Indian Institute of Science, Bangalore
  560012, India}
 
\author{Abhishek K. Singh}
\email {abhishek@mrc.iisc.ernet.in}
\affiliation[Materials Research Centre, Indian Institute of Science, Bangalore
  560012, India]
{Materials Research Centre, Indian Institute of Science, Bangalore
  560012, India}

\title[SnS$_2$ S-M transition]{From semiconductor to metal: A reversible tuning of electronic properties of mono to multilayered SnS$_\text{2}$ under applied strain}
\begin{document}

\begin{abstract}
Controlled variation of the electronic properties of 2D materials by applying strain has emerged as a promising way to design materials for customized applications. Using first principles density functional theory calculations, we show that while the electronic structure and indirect band gap of SnS$_\text{2}$ do not change significantly with the number of layers, they can be reversibly tuned by applying biaxial tensile (BT), biaxial compressive (BC), and normal compressive (NC) strains. Mono to multilayered SnS$_\text{2}$ exhibit a reversible semiconductor to metal transition (S-M) at strain values of 0.17, $-$0.26, and $-$0.24 under BT, BC, and NC strains, respectively. Due to weaker interlayer coupling, the critical strain value required to achieve S-M transition in SnS$_\text{2}$ under NC strain is much higher than for MoS$_\text{2}$. The S-M transition for BT, BC, and NC strains is caused by the interaction between the S-$p_z$ and Sn-$s$, S-$p_x$/$p_y$ and Sn-$s$, and S-$p_z$ and Sn-$s$ orbitals, respectively.
\end{abstract}

\maketitle
\section{Introduction}
The advent of semiconducting two-dimensional (2D) materials such as, MoS$_\text{2}$~\cite{Stephenson2014, Wang2012, Yufei2013, Tonge2012}, phosphorene~\cite{Fei2014}, and SnS$_\text{2}$~\cite{Qu2014, Wei2014, Zhou2013} has attracted immense interest in the scientific community because of their potential applications in electronic, magnetic, optoelectronic, and sensing devices. Modifying their electronic properties in a controlled manner is the key to the success of technologies based on these materials. Some popular techniques used in modifying their properties are electric field, heterostructuring, functionalization, and application of strain. Due to experimental feasibility of strain, it has been a very effective way of transforming and tuning electronic~\cite{Aaditya2015, swasti2012, Hinsche2011, Luo2012, Nayak2015, Guo2013, Nayak2013, kang2015, Vitor2009}, mechanical~\cite{casillas2014} as well as magnetic~\cite{yun2015} properties of these materials. Recent studies have shown its applicability in modifying some of the fundamental properties of bilayer phosphorene and transition metal dichalcogenides (TMDs) where a complete semiconductor to metal (S-M) transition was observed under the applied strain~\cite{Aaditya2015, swasti2012, Hinsche2011, Luo2012, Nayak2015, Guo2013, Nayak2013}. The effect of strain has been studied in not only two-dimensional materials, but also, in materials of other dimensionalities. Some examples include direct to indirect band gap transition in multilayered WSe$_\text{2}$~\cite{Desai2014} and drastic changes in the electronic properties of few layers to bulk MoS$_\text{2}$~\cite{swasti2014, Lijun2015}. Its application is not only limited to electronic properties but it can also engineer the vibrational properties~\cite{Nayak2015, Aaditya2015, Nayak2013} of these materials.

Among all the layered 2D materials, semiconducting SnS$_\text{2}$ is more earth-abundant and environment-friendly and demonstrates a wide spectrum of applications. For example, lithium-~\cite{Li2013, Sathish2012, Chang2012, zhong2012} and sodium-ion batteries~\cite{Zhang2015, Yu2014} with anodes constructed from SnS$_\text{2}$ show high capacity, enhanced cyclability, and excellent reversibility. In addition, SnS$_\text{2}$ shows good photocatalytic activity~\cite{lei2009, zhong2012}. Nanosheets made from SnS$_\text{2}$ are considered to be very good for hydrogen generation by photocatalytic water splitting~\cite{Jing2014, Wei2014}. The visible-light photocatalytic activity of reduced graphene oxide has been shown to be enhanced by doping Cu in SnS$_\text{2}$ sheets~\cite{An2014}. Further usage of SnS$_\text{2}$ in photonics has been demonstrated through fabrication of fast-response SnS$_\text{2}$ photodetectors having an order of magnitude faster photocurrent response times~\cite{su2014, zhong2012} than those reported for other layered materials. This makes SnS$_\text{2}$ a suitable candidate for sustainable and ``green" optoelectronics applications. However, most of these applications rely on the inherent band gap of SnS$_\text{2}$ and therefore, its functionality could be further enhanced by tuning its band gap. Here, we explore this possibility for layered SnS$_\text{2}$ as a function of (i) number of layers and (ii) application of different strains such as biaxial tensile (BT), biaxial compressive (BC), and normal compressive (NC) strains. Band gap tuning in layered materials such as MoS$_\text{2}$~\cite{han2011, swasti2014} and phosphorene~\cite{tran2014, Aaditya2015} has been achieved using these strategies. Interestingly, at zero strain, on increasing the layer number in SnS$_\text{2}$, the band gap was found to be indirect and insensitive to the number of the layers due to the weaker interlayer coupling in SnS$_\text{2}$ compared to MoS$_\text{2}$. In addition, applying strain does not change the nature of the band gap. On the contrary, irrespective of the type of applied strain, a reversible semiconductor to metal (S-M) transition was observed. The S-M transition in bilayer (2L) SnS$_\text{2}$ was achieved at strain values of 0.17, $-$0.26, and $-$0.24, for BT, BC, and NC strains, respectively. The strain values required to achieve this S-M transition in SnS$_\text{2}$ are higher in magnitude compared to MoS$_\text{2}$, which is attributed to the weak interlayer coupling in SnS$_\text{2}$.

\section{Methodology}
The calculations were performed using first-principles density functional theory (DFT)~\cite{Kohn1965} as implemented in the Vienna \textit{ab initio} simulation package (VASP)~\cite{Kress1996, Kresse1996}. Projector augmented wave (PAW)~\cite{Bloch1994, KresseG1999} pseudopotentials were used to represent the electron-ion interactions. In order to obtain an accurate band gap, the Heyd-Scuseria-Ernzerhof (HSE06) hybrid functional~\cite{Heyd2005,Janesko2009,Ellis2011} was used, which models the short-range exchange energy of the electrons by fractions of Fock and Perdew-Burke-Ernzerhof (PBE) exchange~\cite{Kress1996}. The addition of the Fock exchange partially removes the PBE self-interaction which resolves the band gap overestimation problem. Sufficient vacuum was added along the $z$-axis to avoid spurious interactions between the sheet and its periodic images. To model the van der Waals interactions between the SnS$_\text{2}$ layers, we incorporated a semi-empirical dispersion potential ($D$) to the conventional Kohn-Sham DFT energy, through a pair-wise force field through Grimme's DFT-D3 method~\cite{Grimme1799}. The Brillouin zone was sampled by a well-converged 9$\times$9$\times$1 Monkhorst-Pack~\cite{Monkhorst1976} \textbf{k}-grid. Structural relaxation was performed using the conjugate-gradient method until the absolute values of the components of the Hellman-Feynman forces were converged to within 0.005 eV/\AA.

\section{Results and discussion}
\begin{figure*}[!ht]
\centering
\includegraphics[width=\textwidth]{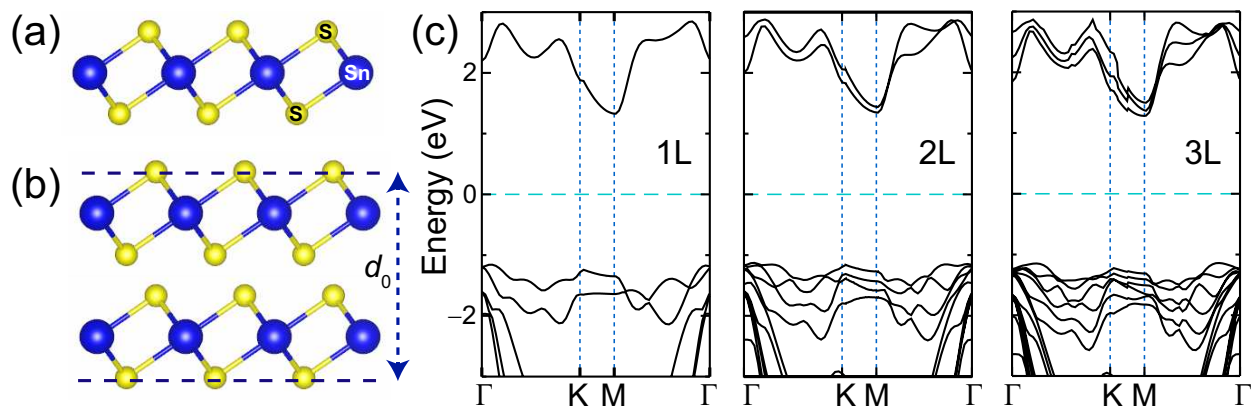}
\caption{Structures of unstrained (a) monolayer SnS$_\text{2}$ and (b) multilayered SnS$_\text{2}$ of equilibrium layer thickness $d_0$. (c) Band structure of unstrained one- (1L), two- (2L) and three-layer (3L) SnS$_\text{2}$. Fermi level is set to 0.}  
\label{Fig:1}
\end{figure*}

SnS$_\text{2}$ belongs to the group-IV dichalcogenide MX$_\text{2}$ family~\cite{Lokha1990}, where, M and X are group IV and chalcogen elements, respectively. The hexagonal SnS$_\text{2}$ monolayer has a CdI$_\text{2}$ layered crystal structure with a space group symmetry of $P\text{3}m\text{1}$~\cite{robert1978} unlike MoS$_\text{2}$, which is a transition metal dichalcogenide with a space group symmetry of $P\text{6}_\text{3}/mmc$. It consists of three atomic sublayers with Sn atoms in the middle layer covalently bonded to S atoms located at the top and bottom layers (Fig.~\ref{Fig:1}(a)). In multilayered SnS$_\text{2}$, the interlayer interactions are van der Waals type~\cite{Patil1971, Whitehouse1979, Mitchell1982, Alamy1977}. The optimized lattice parameters of bulk and monolayer SnS$_\text{2}$ are $a_\text{bulk}=$ 3.689~\AA, $c_\text{bulk}=$ 5.98~\AA, and $a_\text{mono}=$ 3.69~\AA, respectively, which are in good agreement with the experimental values\cite{SunY2012, Toh2013}. SnS$_\text{2}$ prefers AA stacking in the multilayered structure (Fig.~\ref{Fig:1}(b)) and the in-plane lattice parameter $a$ and interlayer distances remain the same as in bulk.

Bulk SnS$_\text{2}$ is an indirect band gap semiconductor with an HSE06 band gap of 2.27 eV. The valence band maxima (VBM) is in between the $\Gamma$ and M points, while the conduction band minima (CBM) is at the L point. The band structures of 1, 2, 3, 4, and 5 layered SnS$_\text{2}$ are calculated and are shown for 1-3 layers in Fig.~\ref{Fig:1}(c). The band structures of all these multilayers show that the number of bands forming VBM and CBM increase with layer number and are equal to the number of layers, as shown in Fig.~\ref{Fig:1}(c). This gives an indication of weak interlayer interaction. Like bulk, all these multilayered SnS$_\text{2}$ structures are indirect band gap semiconductors. The HSE06 band gap values of 1, 2, 3, 4, and 5 layers are 2.47 eV, 2.46 eV, 2.40 eV, 2.35 eV, and 2.34 eV, respectively. The overall variation in the band gap from monolayer to bulk is very small (0.20 eV) demonstrating insensitivity of the electronic structure to the number of layers. It is well-known that experimental growth of these layered materials with a precise control over (a) the number of layers and (b) the quantity of material of a pre-decided thickness, is a difficult task. Therefore, although this insensitivity will limit its application in optical and sensing devices, it can still be beneficial from the nanoelectronics perspective. The devices fabricated from SnS$_\text{2}$ may have minimum noise or error in terms of the change in (a) the contact resistance and (b) carrier mobilities, both arising from the difference in the number of layers, unlike for MoS$_\text{2}$ field effect transistors (FETs)~\cite{krazy2014}.

\begin{figure*}[!h]
\centering
\includegraphics[width=\textwidth]{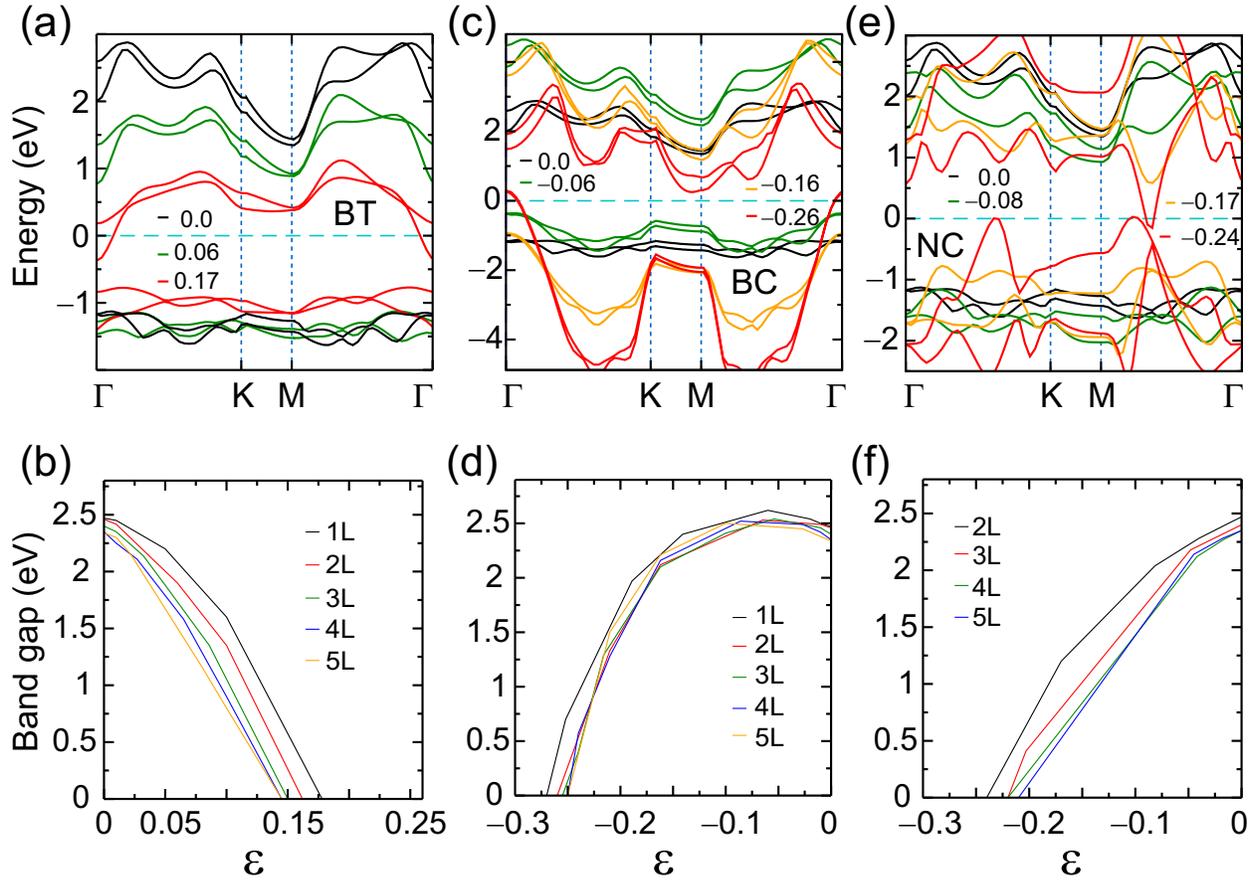}
\caption{Band structures of bilayer SnS$_\text{2}$ as a function of (a) BT, (c) BC, and (e) NC strains. Fermi level is set to 0. Variation of band gap from 1L-to 5L-SnS$_\text{2}$ as a function of (b) BT, (d) BC, and (f) NC strains.}
\label{Fig2}
\end{figure*}

Having shown the insensitivity of the band gap towards the number of layers, we now explore the possibility of electronic structure tuning by applying strain. For any strain value, the nature of the band gap (i.e. indirect band gap) was found to be independent of the number of layers. Therefore, here, we will discuss only the results of bilayer (2L) SnS$_\text{2}$ in detail. First, we study the effect of in-plane uniform BC and BT strains on the electronic structure of 2L-SnS$_\text{2}$. The in-plane strain ($\epsilon$) is given by $\epsilon=\frac{a-a_0}{a_0}$, where, $a$ and $a_0$ are lattice parameters with and without strain, respectively. For the unstrained 2L-SnS$_\text{2}$, the VBM is in between the $\Gamma$ and M points, while the CBM is at the M point. The conduction bands CBM and CBM$-$1 are dispersive at the M point, whereas, the valence bands VBM and VBM$-$1, are not very dispersive. Hence, the overall curvature as well as the mobilities of these valence bands are expected to be quite low. With increasing BT strain, the VBM and VBM$-$1 remain flat and are still located in the $\Gamma$-M region as shown in Fig.~\ref{Fig2}(a). At a strain value of 0.06, the CBM shifts to the $\Gamma$ point while still preserving the indirect nature of the band gap. In addition, the curvature of the conduction bands at the M ($\Gamma$) point decreases (increases), implying that, locally, the mobility of $n$-type carriers at this point may decrease (increase). However, the overall mobility of the $n$-type carriers may not change significantly, owing to the compensation of these local changes in the mobilities. Also, irrespective of the strain, the shape of the conduction as well as the valence bands do not change. At a critical strain value of 0.17, the CBM at $\Gamma$ crosses the Fermi level (Fig.~\ref{Fig2}(a)), rendering 2L-SnS$_\text{2}$ metallic (Fig.~\ref{Fig2}(b)).

Under BC strain, similar changes in the band dispersion of 2L-SnS$_\text{2}$ are observed. Surprisingly, unlike the BT strain, at a strain value of $-$0.06, there is a significant shift in the CBM away from the Fermi level as compared to the VBM resulting in an increase in the band gap. However, irrespective of strain, the CBM remains at the M point. The VBM remains at the $\Gamma$-M location upto a strain of $-$0.06, beyond which, it moves to the $\Gamma$ point. In addition, both, the valence as well as conduction bands become more dispersive with strain, indicating that the overall mobilities of $p$- and $n$-type carriers may increase. Similar to BT strain, the movement of VBM and CBM towards the Fermi level leads to the band gap reduction, while still preserving an indirect band gap, as shown in Fig.~\ref{Fig2}(c). At a critical strain of $-$0.26, the S-M transition occurs (Fig.~\ref{Fig2}(d)), with the VBM crossing the Fermi level (Fig.~\ref{Fig2}(c)).

We next study the changes in band structure of SnS$_\text{2}$ under normal compressive (NC) strain ($\epsilon_\text{NC}$) defined as $\epsilon_\text{NC}=\frac{d-d_0}{d_0}$, where $d$ and $d_0$ (Fig.~\ref{Fig:1}(b)) are the layer thickness at applied and zero strains. The minimum required number of layers for NC strain is two. The NC strain was applied perpendicular to the plane of the multi-layers of SnS$_\text{2}$. Constrained relaxation mechanism was incorporated wherein the atoms in upper and lower layers were restricted to move along the normal direction at each strain value. With increasing strain, the band gap reduces smoothly and becomes metallic at $-$0.24, as shown in Figs.~\ref{Fig2}(e) and (f). In the strain range of $-$0.08 to $-$0.17, the VBM remains in between the $\Gamma$ and M points, whereas there is a drastic shift in the position of CBM from M to in between $\Gamma$ and M points. Similar to BC strain, the bands become more dispersive with increasing NC strain, indicating that the overall mobilities of $p$- and $n$-type carriers can increase. At $-$0.24 strain, a band crossing occurs at the Fermi level rendering it metallic.

\begin{figure*}[!ht]
\centering
\includegraphics[width=\textwidth]{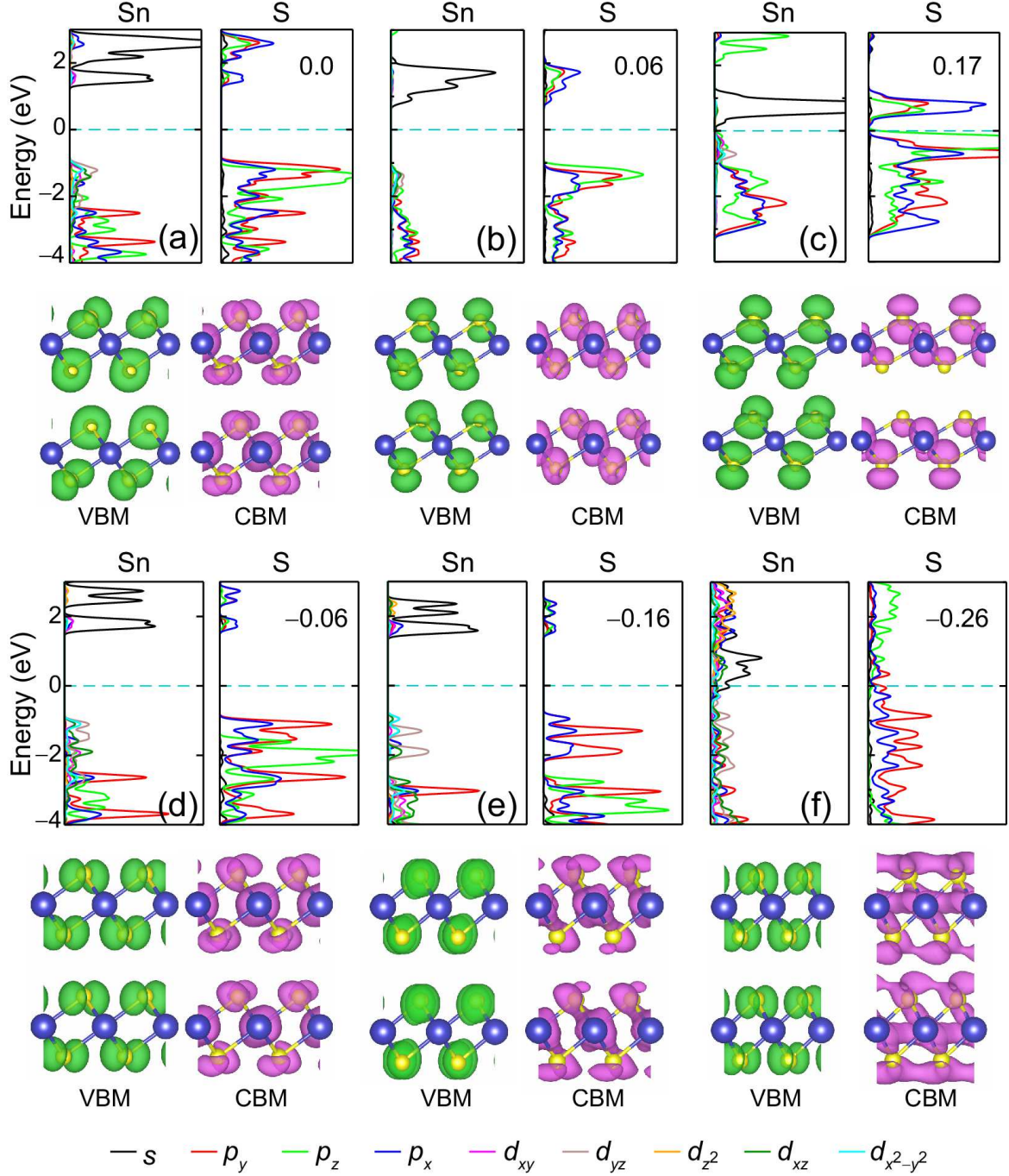}
\caption{(a)-(f) LDOS and corresponding band-decomposed charge densities (VBM and CBM) for 2L-SnS$_\text{2}$ under BT and BC strains. The isosurface has been set to 0.006 e/\AA$^\text{3}$.}
\label{Fig:3}
\end{figure*}

Next, we discuss the mechanism of S-M transition under BT, BC, and NC strains. We calculated the angular momentum projected density of states (LDOS) and band-decomposed charge density of 2L-SnS$_\text{2}$ as a function of these strains. In unstrained SnS$_\text{2}$, the valence and conduction bands originate predominantly from S-$p_x$/$p_y$/$p_z$, and the direction-independent Sn-$s$ orbitals, respectively, as shown in Fig.~\ref{Fig:3}(a). With increasing BT strain up to 0.06, S-$p_z$ crosses the S-$p_x$ and S-$p_y$ orbitals, and becomes the frontier orbital. The band gap begins to reduce due to the movement of these frontier orbitals (Sn-$s$ and S-$p_z$) towards the Fermi level. For strains beyond 0.06, the intralayer interactions between these orbitals begin to increase. This is also confirmed by the band-decomposed charge density plots (Fig.~\ref{Fig:3}(b)). At a critical strain of 0.17, there is a strong hybridization between these orbitals, as is also observed in the CBM band-decomposed charge density (Fig.~\ref{Fig:3}(c)), which finally leads to the semiconductor to metal transition.

In the case of BC strain, the S-$p_z$ orbital crosses the tails of the S-$p_x$ and S-$p_y$ orbitals and moves away from the Fermi level, as shown in Fig.~\ref{Fig:3}(d). Hence, the S-$p_x$/$p_y$ and Sn-$s$ orbitals contribute predominantly to the valence and the conduction bands, respectively. With a strain upto $-$0.06, the band gap initially increases, as observed from the band structure in Fig.~\ref{Fig2}(d). This increase is due to the slight upward (downward) shift in the Sn-$s$ (S-$p_x$/$p_y$) orbitals along with the crossing of the S-$p_z$ orbital, causing the conduction (valence) bands to move up (down) as shown in Fig.~\ref{Fig:3}(d). A similar band gap increase at an intermediate BC strain is observed in MoS$_\text{2}$~\cite{swasti2014}. As shown in the VBM band-decomposed charge density in Fig.~\ref{Fig:3}(d), the wavefunction spread between the S atoms (above and below Sn) in a layer decreases compared to the unstrained case (Fig.~\ref{Fig:3}(a)) due to the shifting of the valence orbitals $p_x$, $p_y$, and $p_z$ of the S atom. This localization causes the increase in the band gap. As we go beyond $-$0.06, the band gap begins to decrease because of the movement of the S-$p_x$/$p_y$ (in-plane orbitals) and Sn-$s$ orbitals towards the Fermi level. At a critical strain of $-$0.26, this movement results in S-M transition. Here, the in-plane orbitals contribute to S-M transition, whereas in BT strain, in addition to the in-plane orbitals, out-of-plane orbitals also play a role in triggering S-M transition.

\begin{figure*}[!ht]
\centering
\includegraphics[width=\textwidth]{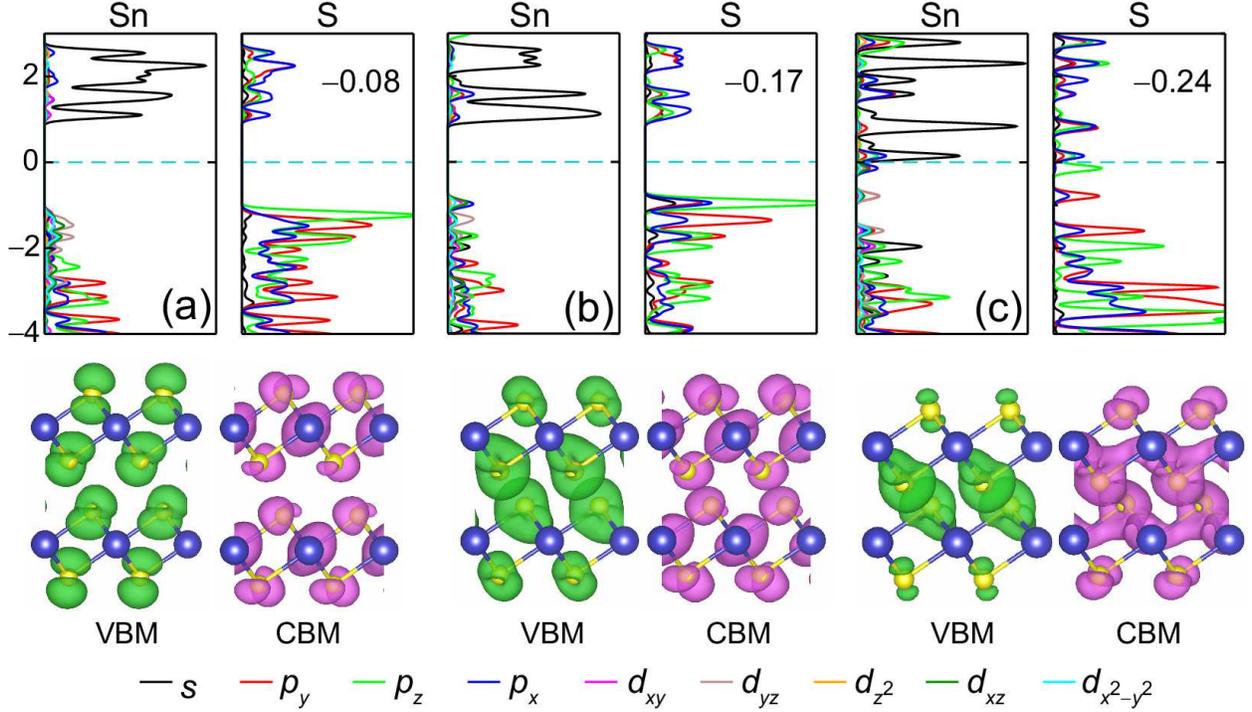}
\caption{(a)-(c) LDOS and corresponding band-decomposed charge densities (VBM and CBM) for 2L-SnS$_\text{2}$ under NC strain. The isosurface  has been set to 0.006 e/\AA$^\text{3}$.}
\label{Fig:4}
\end{figure*}

With increasing NC strain up to $-$0.17, the Sn-$s$ and S-$p_x$/$p_y$/$p_z$ orbitals begin to move towards the Fermi level, as shown in Figs.~\ref{Fig:4}(a) and (b). For strains beyond $-$0.17, these orbitals start overlapping because of the reduction in the interlayer distance. At a critical strain of $-$0.24, interlayer interactions become enhanced and as a consequence, results in strong hybridization between S-$p_z$ and Sn-$s$ orbitals (Fig.~\ref{Fig:4}(c)). These orbitals cross the Fermi level and lead to the S-M transition. From the band-decomposed charge analysis, at this critical strain, the presence of out-of-plane lobes ($p_z$) in the VBM and $s$ orbital in CBM are mainly responsible for S-M transition, as shown in Fig.~\ref{Fig:4}(c).

Among the applied strains, BT strain results in the fastest S-M transition. Although the critical strain required in the case of BT strain is the lowest ($|\epsilon|=$ 0.17 compared to $|\epsilon|=$ 0.26 and 0.24 in the case of BC and NC strains, respectively), the NC strain is more feasible compared to the biaxial strains. This is because the intralayer bonding between the atoms is much stronger compared to the weaker interlayer van der Waals interactions.

\begin{figure*}[!ht]
\centering
\includegraphics[width=\textwidth]{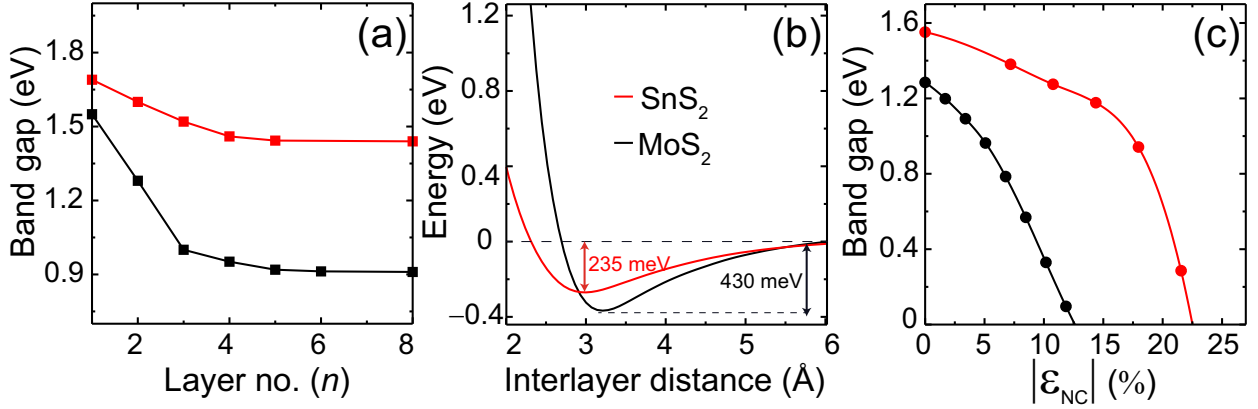}
\caption{(a) Band gap variation as a function of number of layers, (b) interlayer coupling energy for the corresponding interlayer distance under NC strain, and (c) band gap as a function of NC strain $\%$, for both SnS$_\text{2}$ (red line) and MoS$_\text{2}$ (black line).}
\label{Fig:5}
\end{figure*}

\begin{figure*}[!ht]
\centering
\includegraphics[width=\columnwidth]{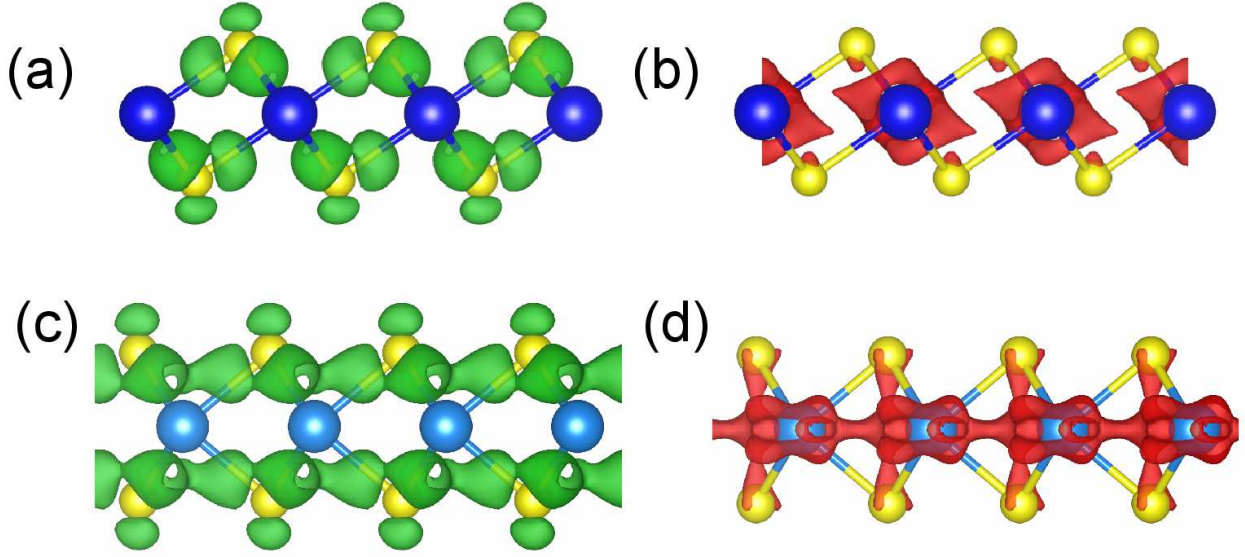}
\caption{Charge accumulation (light-green) and depletion (red) of (a)-(b) SnS$_\text{2}$ and (c)-(d) MoS$_\text{2}$. The isosurface is set to 0.006 e/\AA$^\text{3}$.}
\label{fig:6}
\end{figure*}

Having explored band gap tuning using BT, BC, and NC strains, it is interesting to perform a comparative study of the band gap variation as a function of (a) interlayer coupling between the layers and (b) applied NC strain, in both 2L-SnS$_\text{2}$ and 2L-MoS$_\text{2}$. Here, we use the generalized gradient approximation (GGA) of the PBE form to model the exchange-correlation. Fig.~\ref{Fig:5}(a) shows the band gap as a function of the number of layers for the unstrained structures of MoS$_\text{2}$ and SnS$_\text{2}$. As we go from 1L to 5L, the change in the band gap is small for SnS$_\text{2}$ ($\sim$ 0.24 eV) compared to that in MoS$_\text{2}$ ($\sim$ 0.59 eV), indicating that the band gap of SnS$_\text{2}$ is less sensitive to the number of layers. The reason for this insensitivity is attributed to the weak interlayer coupling in SnS$_\text{2}$ compared to MoS$_\text{2}$. Fig.~\ref{Fig:5}(b) shows the interlayer coupling energy as a function of interlayer distance for 2L-SnS$_\text{2}$ and 2L-MoS$_\text{2}$. Here, the adjacent layers in SnS$_\text{2}$ are weakly coupled to each other, with a coupling energy of 235 meV/unit cell, relative to that of MoS$_\text{2}$ (430 meV/unit cell). The weaker interlayer coupling in SnS$_\text{2}$ can be explained based on the higher Sn electron affinity ($\chi_\text{Sn}= $~107.3 kJ/mol)~\cite{Sn_EA} compared to Mo electron affinity ($\chi_\text{Mo}= $~72.3 kJ/mol)~\cite{Mo_EA}. A higher $\chi$ value indicates a lower tendency to interact with the neighboring atoms. Therefore, the polarizability of the Sn atom is much lower than in Mo causing a weaker interlayer coupling in SnS$_\text{2}$. The above observation is also confirmed by the charge accumulation and depletion calculations of 1L-SnS$_\text{2}$ and 1L-MoS$_\text{2}$ as shown in Figs.~\ref{fig:6}(a)-(d) wherein, the lower accumulation of charge around the S atoms and lower depletion of charge around Sn atoms indicates weaker polarizability.
As a consequence, the out-of-plane linear elastic modulus ($C_z$), i.e.
\begin{equation}
C_z=\frac{\partial^\text{2}E_\text{coupling}}{\partial z^\text{2}}
\end{equation}
where, $z$ is the interlayer distance and $E_\text{coupling}$ is the interlayer coupling energy, of SnS$_\text{2}$ ($C_z= $~11.1 GPa) is nearly half of that of MoS$_\text{2}$ ($C_z= $~20.6 GPa). As a function of NC strain, in bilayer MoS$_\text{2}$, due to stronger interlayer coupling, the strain dependence of band gap is more prominent (0.11 eV/unit strain) than SnS$_\text{2}$ (0.026 eV/unit strain), as shown in Fig.~\ref{Fig:5}(c). Therefore, the overlap of wave functions within the adjacent layers is much stronger in MoS$_\text{2}$ than in SnS$_\text{2}$, and hence a small change in the interlayer distance can lead to a band renormalization in MoS$_\text{2}$. For the same reason, semiconductor to metal transition is observed at a relatively high strain 23$\%$ for 2L-SnS$_\text{2}$ compared to 2L-MoS$_\text{2}$ (12$\%$) under NC strain (Fig.~\ref{Fig:5}(c)). Therefore, the insensitivity of the indirect band gap of SnS$_\text{2}$ towards the number of layers and less dependency on strain can be attributed to the weak coupling between the layers.
 
In conclusion, using density functional theory based calculations, we have shown the tuning of the band gap as well as reversibility in S-M transition for multilayers of SnS$_\text{2}$ under different strains (BT, BC, and NC). With increasing strain, there is a smooth reduction in the band gap for all the strains, with a small increase observed at $-$0.06 BC strain. The critical strain at which S-M transition is achieved, decreases significantly, as we go from mono to five layers for all the strains. Moreover, in order to understand the S-M transition under different strains, we analyzed the contribution from different molecular orbitals by performing LDOS and band-decomposed charge density calculations. For BT, BC, and NC strains, the interaction between the S-$p_z$ and Sn-$s$, S-$p_x$/$p_y$ and Sn-$s$, and S-$p_z$ and Sn-$s$ orbitals, respectively, trigger S-M transition. For comparison, we studied the band gap trends in 2L-SnS$_\text{2}$ and 2L-MoS$_\text{2}$. At zero strain, the change in the band gap of SnS$_\text{2}$ is $\sim$ 0.24 eV, which is more than half of that in MoS$_\text{2}$ ($\sim$ 0.59 eV), while going from monolayer to five layer. Moreover, the band gap in SnS$_\text{2}$ remains indirect irrespective of the number of layers as well as applied strain. The insensitivity of the band gap in SnS$_\text{2}$ to the number of layers is attributed to the weaker interlayer coupling (235 meV/unit cell) than in MoS$_\text{2}$ (430 meV/unit cell) due to lower polarizability. In terms of band gap dependence under NC strain, SnS$_\text{2}$ is less sensitive to strain (0.026 eV per unit strain) than MoS$_\text{2}$ (0.11 eV per unit strain).

\section{Acknowledgements}
The authors thank Tribhuwan Pandey for fruitful discussions, and the Supercomputer Education and Research Centre, IISc, for providing the required computational facilities for the above work. The authors acknowledge DST Nanomission for financial support. 

\providecommand{\latin}[1]{#1}
\providecommand*\mcitethebibliography{\thebibliography}
\csname @ifundefined\endcsname{endmcitethebibliography}
  {\let\endmcitethebibliography\endthebibliography}{}

\end{document}